\documentclass[prl,twocolumn,amsmath,amssymb,longbibliography,floatfix]{revtex4-2}
\usepackage[T1]{fontenc}
\usepackage[latin9]{inputenc}
\usepackage{epsfig, graphicx,graphics,float}
\usepackage{mathrsfs}

\usepackage{appendix}
\usepackage{amscd}
\usepackage{bm}
\usepackage{psfrag}
\usepackage{bbm} 
\usepackage{babel}
\usepackage{wasysym }
\usepackage{mathrsfs}
\usepackage{color}
\usepackage[breaklinks]{hyperref}
\hypersetup{colorlinks,linkcolor={magenta},citecolor={blue},urlcolor={blue}}  

\def\be{\begin{equation}}
\def\ee{\end{equation}}
\def\bea{\begin{eqnarray}}
\def\eea{\end{eqnarray}}
\def\bmat{\begin{pmatrix}}
\def\emat{\end{pmatrix}}
\def\bs{\begin{split}}
\def\es{\end{split}}

\def\~{$\approx$}
\def\bra{\langle}
\def\ket{\rangle}
\def\up{\uparrow}
\def\dw{\downarrow}
\def\dag{\dagger}

\begin{document}

\title{Intrinsic non-magnetic \texorpdfstring{$\phi_0$}{TEXT} Josephson junctions in twisted bilayer graphene}

\author{M. Alvarado, P. Burset and A. Levy Yeyati}
\affiliation{
Departamento de F{\'i}sica Te{\'o}rica de la Materia Condensada, Condensed Matter Physics Center (IFIMAC)
and Instituto Nicol{\'a}s Cabrera, 
Universidad Aut{\'o}noma de Madrid, 28049 Madrid, Spain}

\begin{abstract}
Recent experiments have demonstrated the possibility to design highly controllable junctions on magic angle twisted bilayer graphene, enabling the test of its superconducting transport properties. We show that the presence of chiral pairing in such devices manifests in the appearance of an anomalous Josephson effect ($\phi_0$ behavior) even in the case of symmetric junctions and without requiring any magnetic materials or fields. 
Such behavior arises from the combination of chiral pairing and nontrivial topology of the twisted bilayer graphene band structure that can effectively break inversion symmetry. Moreover, we show that the $\phi_0$ effect could be experimentally enhanced and controlled by electrostatic tuning of the junction transmission properties. 
\end{abstract}

\maketitle

{\it Introduction.---} Since its discovery in 1962~\cite{Josephson}, the Josephson effect has emerged as one of the most remarkable manifestations of quantum coherence at the macroscopic scale with multiple technological applications~\cite{Tafuri}. Material platforms on which Josephson junctions (JJs) can be fabricated do not cease to grow, ranging from conventional or unconventional superconductor tunnel junctions~\cite{review-HTC} to hybrid nanostructures including exotic materials~\cite{DeFranceschi2010}. 
The discovery of superconducting phases in magic angle twisted bilayer graphene (MATBG)~\cite{Cao2018} provides an additional playground for the study of the Josephson effect. In fact, the possibility to produce tunable junctions on this material by electrostatic gating has been recently demonstrated~\cite{Rodan-Legrain2021,deVries2021}, with enabled phase control in ring shaped configurations \cite{portoles2022}, and JJs exhibiting unconventional Fraunhofer patterns have been found~\cite{Diez-Merida}. 

This work is in parallel with worldwide efforts to clarify the origin and precise nature of unconventional superconductivity in MATBG~\cite{Cao2018,Thomale2020,oh2021, shavit2021, kim2022, di2022, lake2022, khalaf2022}. Many theoretical studies point to the possibility of chiral ($d+id$ or $p+ip$) pairing symmetry arising as a result of electron correlations and the peculiar topology of flat bands in MATBG~\cite{d+id-1,d+id-2,d+id-3,d+id-4,d+id-5,d+id-6}. Furthermore, robust nematic behaviour was observed for a variety of twist angles in the superconducting phase of twisted bilayer systems~\cite{Nadj_Perge2019, kerelsky2019, jiang2019, cao2021},
 a phenomenon that has attracted recent theoretical interest~\cite{julku2020, Chichinadze2020, fernandes2020, yu2021, park2021, lothman2022}. 
The next key step in the field is thus to find reliable signatures that would help us to elucidate the pairing mechanism in MATBG. 

In this Letter we address this issue showing that chiral pairing symmetry would manifest in an anomalous $\phi_0$ behavior (a nonzero supercurrent in the absence of superconducting phase difference) in  monolithic MATBG JJs. We demonstrate that this effect is a consequence of the nontrivial topology of the normal MATBG bands and thus cannot be captured by trivial models. We further show that the effect is enhanced for extended junctions and can be controlled by electrostatic gating of a middle normal region. 

Before entering into the peculiarities of MATBG JJs, let us give a more general context on the topic of $\phi_0$ junctions. In general, $\phi_0$ behavior requires breaking time reversal symmetry (TRS) and inversion symmetry, having been predicted to appear in different systems by the combined effect of spin-orbit interactions and magnetic fields~\cite{Reynoso,Zazunov,Tkachov_2015,Dolcini,Bergeret2015,yuan2022,he2022,Tanaka22,Lu_2022} or in junctions through magnetic metals with broken inversion symmetry~\cite{Buzdin,Black-Schaffer,Kopasov_2021,Vasenko_2022}. In all these proposals the broken symmetries are linked to the spin degree of freedom and we could refer to them as ``magnetic''-$\phi_0$ junctions. 

In the case of MATBG the valley degree of freedom comes into play, providing new possibilities. In Ref.~\onlinecite{Xie} it has been suggested that $\phi_0$ behavior could appear in MATBG JJs with conventional $s$-wave pairing provided that the junction is established through a ``valley polarized'' region. This mechanism has also been proposed to explain recent observations of anomalous Josephson effect in twisted trilayer JJs~\cite{Lin2022}. By contrast, 
we here analyze direct junctions between two MATBG regions with chiral pairing. An important aspect of MATBG which guides our study is that the large size of its moir\'e pattern (>10 nm) would allow transport experiments on junctions along well defined directions on the moir\'e lattice. 
We show that, only when the nontrivial topology of the MATBG is taken into account, $\phi_0$ behavior can appear spontaneously when the junction is defined along certain directions on the moir\'e pattern that break translation invariance, or along any orientation when graphene's sublattice symmetry is broken. We argue that these effects would allow one to distinguish among possible pairing symmetries in MATBG.

\begin{figure*}[t]
\centering
\includegraphics[width=1\textwidth]{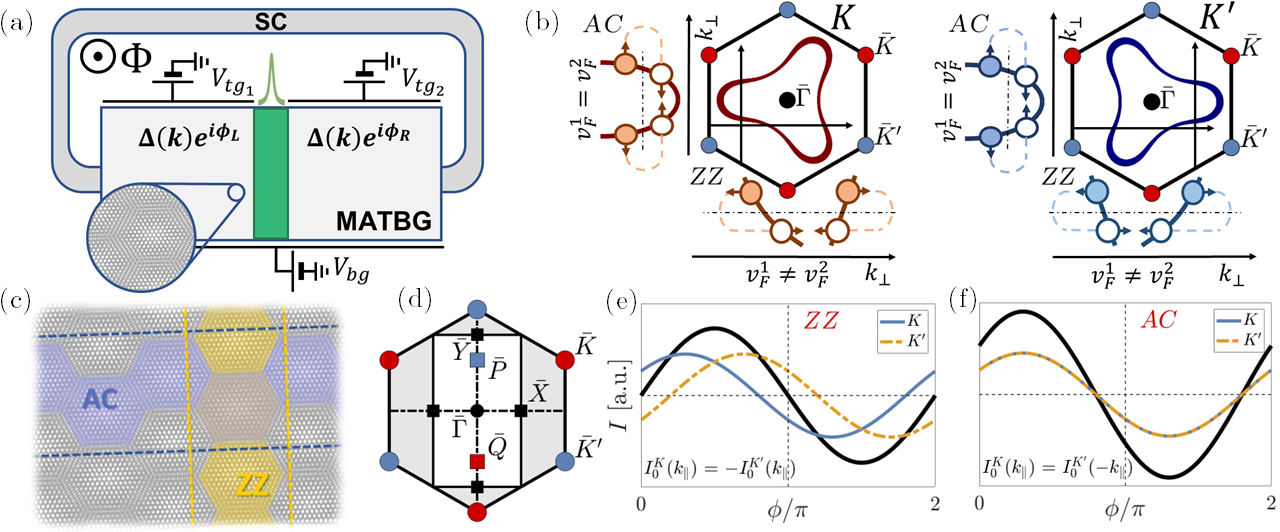}
\caption{(a) Scheme of a MATBG Josephson junction: side gates control the doping level on the two sides and the middle barrier height. The junction is immersed in a loop that determines the superconducting phase difference $\phi=\phi_L-\phi_R$. 
 (b) Typical normal-state Fermi surfaces at the $K$ and $K'$ valleys in the hexagonal Brillouin zones. The Andreev reflection coefficients on opposite Fermi points for a given $k_{\parallel}$ can differ either due to trigonal warping, leading to $v^1_F\ne v^2_F$ (ZZ), or due to nontrivial topology combined with chiral pairing (AC). 
 (c) Representation of AC and ZZ directions on the moir\'e lattice. 
 (d) Hexagonal and folded rectangular Brillouin zones indicating high symmetry points. 
 (e,f) Schematic CPRs at the two valleys for the AC and ZZ cases. 
}
\label{fig1}
\end{figure*}

{\it Qualitative description.---} 
We consider a junction defined on a bulk MATBG sample as schematically illustrated in Fig.~\ref{fig1}(a). External gates can be used to tune the doping level on the two sides of the junction independently and each side is characterized by the presence of a generic chiral pairing. The barrier height, and therefore the junction transmission, can also be externally controlled. The junction is placed on a loop that enables control over the superconducting phase difference $\phi$, determining the junction current-phase relation (CPR), $I_J(\phi)$. 

At band fillings close to the van Hove singularities (VHSs), where superconductivity
is observed, MATBG is characterized by a normal Fermi surface exhibiting trigonal warping~\cite{Bistritzer2011,Koshino2018}. Typical shapes of these surfaces on the two valleys (denoted by $K$ and ${K^\prime}$) are shown in Fig.~\ref{fig1}(b). 
The main effect of trigonal warping is to break the intra-valley inversion symmetry along certain directions, i.e., $E(k_{\perp}) \ne E(-k_{\perp})$, where $k_{\perp}$ denotes the wavevectors perpendicular to the junction interface. The Fermi velocities $v_F$ at the Fermi points for a given parallel momentum $k_{\parallel}$ thus differ. 
This is the case, for instance, for the junctions defined parallel to the line joining the moir\'e minivalleys $\bar{K}-\bar{K^\prime}$ (i.e., $k_{\parallel} = k_y$), which we call ``zigzag'' (ZZ) in analogy with pristine graphene. On the contrary, junctions defined perpendicular to the $\bar{K}-\bar{K^\prime}$ line (i.e., $k_{\parallel} = k_x$), called ``armchair'' (AC), are not affected by trigonal warping in the normal state~\cite{sainz2022}. However, as we discuss below, such symmetry can be broken even for AC junctions due to the combined effect of normal state topology and pairing properties. As illustrated in Fig.~\ref{fig1}(c), these junction orientations correspond in real space to lines joining nearest neighbors (ZZ) or next nearest neighbors (AC) ``sites" in the moir\'e pattern, which can be associated with the charge centers. 

From this definition we observe that an AC junction breaks the moir\'e translation invariance along the interface. Periodicity along the junction interface (i.e., along $k_{\parallel}=k_x$) is recovered by cell duplication, which corresponds to folding the Brillouin zone into a rectangular one as shown in Fig.~\ref{fig1}(d). Consequently, in the superconducting state the Andreev bound states (ABSs) formed at such interface result from coupling scattering states of different type on both sides, which might differ by a topological phase. 
While this asymmetry is a necessary condition, to obtain a net $\phi_0$ behavior this phase difference should survive after $k_{\parallel}$ and valley integration.   

We find that spontaneous valley supercurrents $I^K_J(\phi=0)\neq 0$ appear in the AC case due to the combined effect of chiral pairing and nontrivial band topology, and in the ZZ due to Fermi-surface warping. 
However, while the ZZ currents on opposite valleys cancel each other [i.e., $I^K_J(0)=-I^{K^\prime}_J(0)$, see Fig.~\ref{fig1}(e)], for AC we have $I^K_J(0)=I^{K^\prime}_J(0)\neq 0$. 
Consequently, the valley currents do not compensate leading to a net $\phi_0$-junction behavior, see Fig.~\ref{fig1}(f).  
As we show below, breaking graphene's sublattice symmetry yields an anomalous CPR also for junctions defined on a ZZ direction only in models where nontrivial band topology is taken into account. 

Our main result, which we expose in detail below, is thus that the presence of chiral pairing on a nontrivial MATBG band manifests as an anomalous Josephson effect. This is always the case in AC junctions, due to the naturally broken moir\'e translation invariance at the interface, and can be induced on ZZ junctions by breaking graphene's sublattice symmetry. 

{\it CPR calculations for MATBG junctions.---} To give a quantitative estimation of the effect discussed above, we consider a six band model (6BM) for MATBG junctions which accounts for both the appropriate flat bands dispersion and their nontrivial (fragile) topology~\cite{Vishwanath2019}. 
Our calculations are based on the recursive Green functions method introduced in Refs.~\onlinecite{Zazunov2016,Alvarado2021,Alvarado2022}, which we extend here to the superconducting case, see the Supplemental Material (SM)~\cite{SM} where we also include the pairing implementation for the trivial two band model (2BM) ~\cite{Koshino2018}.
To introduce chiral superconductivity in these models we project the pairing order parameter onto the directional $p_\pm$ orbitals which provide the main contribution to the flat bands below the VHS~\cite{SM}. For simplicity, we only consider here intervalley, spin-singlet chiral $d$-wave superconductivity with zero center-of-mass momentum. From the irreducible representations of the crystal symmetry group~\cite{pangburn2022} in the triangular lattice we build up the form factors for the pairing
\bea
\Delta_{d_{x^2-y^2}} &=& \frac{\Delta}{\sqrt{3}} \left[ \cos{(k_y L_y)}-\cos{(k_x L_x/2)}\cos{(k_y L_y/2)} \right], \nonumber \\
\Delta_{d_{xy}} &=& \Delta\sin{(k_x L_x/2)}\sin{(k_y L_y/2)},
\eea 
where the chiral pairing is the superposition $\Delta_{d+id'} = \Delta_{d_{x^2-y^2}}+i\Delta_{d_{xy}}$, $L_{x,y}$ are the orthogonal lattice vectors in the doubled unit cell, and we set the value of $\Delta$ as roughly one order of magnitude smaller than the bandwidth. Some other configurations (e.g., chiral $p$-wave with $S^z=0$ or superpositions of nematic chiral orders with local $s$-wave) could be implemented using the same approach, but have minor effects on the CPR. 
Notice that chiral superconductivity is a necessary ingredient to obtain $\phi_0$ behavior as it breaks TRS. However, not all TRS-breaking pairings lead to $\phi_0$ effect, e.g., nodal intravalley pairings never show uncompensated valley currents for any junction configuration studied in this work~\cite{SM}. In this sense, $\phi_0$ behaviour is intrinsically linked only to chiral superconductivity.

Along with pairing, we consider the effect of a sublattice symmetry breaking perturbation ($\delta_S$) acting over $p_\pm$ that can be produced by alignment with the \textit{h}BN substrate \cite{SM}. We note that the nontrivial topology of the bands in the 6BM is already reflected in an anisotropic response of the superfluid weight \cite{SM}. The relevant effect of remote bands for superconductivity in MATBG has been discussed in several theoretical works~\cite{peotta2015, liang2017, hu2019, julku2020, torma2021, rossi2021}. 

The junction is defined as a smooth barrier at the moir\'e scale for which $k_\parallel$ is a good quantum number. Its effective transmission is controlled by a coefficient $\tau$ such that for a symmetric junction ($SS$) with the same chemical potential on both leads, $\mu_L = \mu_R = \mu_0$, we recover the bulk system at zero phase bias and $\tau=1$. 
We also consider asymmetric junctions ($SS^\prime$) in which $\mu_L \neq \mu_R$ and symmetric junctions through a finite-sized, heavily-doped normal region ($SNS$) where $|\mu_L| = |\mu_R| \ll |\mu_c|$ and $\tau=1$. 

\begin{figure}[t!]
\includegraphics[width=8.6cm]{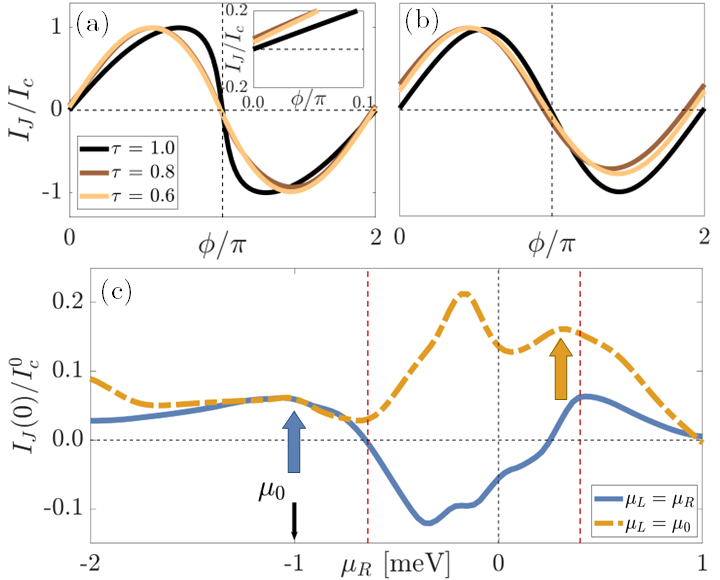}
\caption{(a-b) Current-phase relation for AC junctions in the 6BM for different values of the effective transmission $\tau$. 
(a) Symmetric $SS$ junction with $\mu_L=\mu_R=-1$meV and (b) asymmetric $SS'$ junction with $\mu_L=-1$meV and $\mu_R=0.3$meV. The inset in (a) shows a close up of the $\phi \simeq 0$ region.
(c) Zero-phase current $I_J(0)$ as a function of the chemical potential for symmetric ($\mu_L=\mu_R$) and asymmetric ($\mu_L=\mu_0$, with $\mu_0=-1$meV) junctions. $I_J(0)$ is normalized to the critical current $I_c^0$ in the symmetric case with $\mu_L=\mu_R=\mu_0$ and $\tau = 0.8$. 
Red dashed lines mark the position of the VHSs and color arrows indicate the situation for the CPRs in (a) and (b). 
}
\label{fig2}
\end{figure}

We first discuss the case of AC junctions. Figure~\ref{fig2}(c) shows the evolution of the zero-phase supercurrent $I_J(0)$ for $\tau=0.8$ as a function of the chemical potential, both in the case of equal ($\mu_L=\mu_R$) and unequal ($\mu_L$ fixed, varying $\mu_R$) doping levels for the 6BM. Since the critical currents for all cases are of the same order, we normalize $I_J(0)$ to the critical current $I^0_c$ for $\mu_L=\mu_R=-1$ meV. The red dashed lines indicate the VHS positions. 

While the most striking result is the presence of $\phi_0$ behavior for a symmetric junction without external fields, the computed $|I_J(0)/I^0_c|$ values are relatively small (less than $10\%$) in the range of doping levels where superconductivity in MATBG is expected to appear. 
We observe similar values of $I_J(0)$ for symmetric junctions with p or n doping close to the VHSs, see Fig.~\ref{fig2}(c).
The $\phi_0$ effect is generally larger for asymmetric junctions of n-p type, i.e., with $\mu_L<0$ and $\mu_R>0$. 
In contrast to the symmetric case, in this situation we observe no changes in the sign of $I_J(0)$ as a function of the doping level.
Finally for both symmetric and asymmetric junctions we find large non-vanishing $|I_J(0)/I_c|$ values close to charge neutrality $\mu=0$. We notice, however, that experimental samples show no robust superconducting dome at those fillings and that the superconducting state could have different properties close to charge neutrality~\cite{lu2019,oh2021}. 

\begin{figure}[t!]
\includegraphics[width=8.6cm]{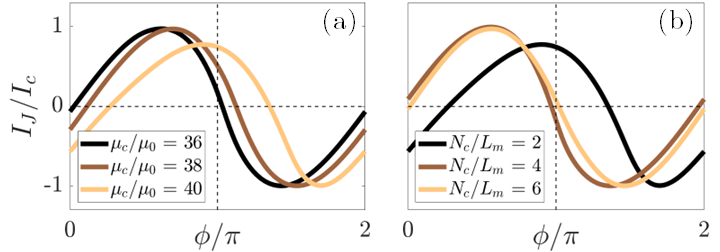}
\caption{Josephson effect for AC $SNS$ junctions described by the 6BM. (a) CPR for different potential barrier heights $\mu_c = -36, -38, -40$ meV, where the length of the normal region is $N_c = 2$ (in units of moir\'e lattice length $L_m$). (b) CPR for several values of the width of the normal region $N_c = 2, 4, 6$ (approximately $25, 55 , 80$ nm). In all cases the chemical potentials are $\mu_L = \mu_R = \mu_0 = -1$ meV and $\tau=1$. 
}
\label{fig3}
\end{figure}

Having demonstrated the anomalous Josephson effect of short symmetric and asymmetric junctions with chiral pairings, we now show that the $\phi_0$ effect can be enhanced and tuned by a middle normal region in the $SNS$ configuration. 
We thus assume symmetric junctions including a finite normal central region with large doping beyond the flat bands, so that the barrier transmission is limited by the chemical potential mismatch. 
We show in Fig.~\ref{fig3} that it is possible to tune the $|I_J(0)/I_c|$ ratio by varying the doping levels (a) or the length (b) of the central region. 
Notice that doping the central region at the dispersive bands or increasing its length reduces the critical current by an order of magnitude with respect to the cases analyzed in Fig.~\ref{fig2}. 

\begin{figure}[t!]
\includegraphics[width=8.6cm]{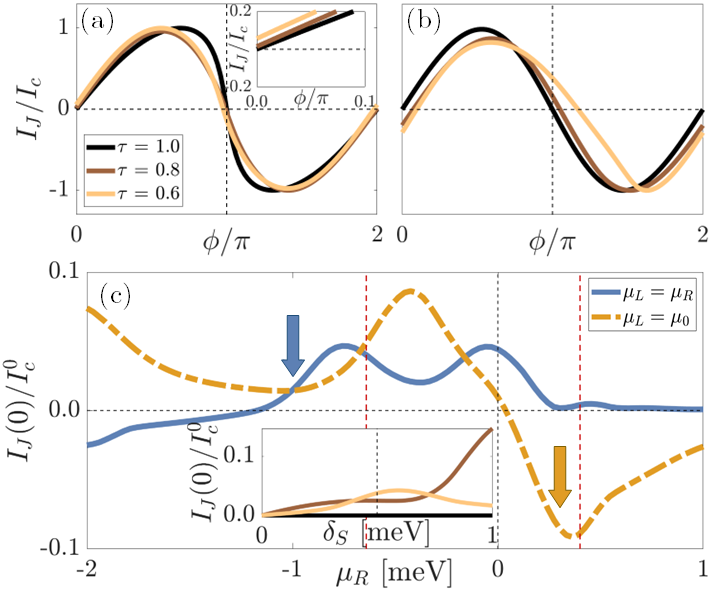}
\caption{(a-b) Current-phase relation for ZZ junctions without sublattice symmetry in the 6BM for different values of the effective transmission $\tau$. 
(a) Symmetric $SS$ junction with $\mu_L=\mu_R=-1$meV and (b) asymmetric $SS'$ junction with $\mu_L=-1$meV and $\mu_R=0.3$meV. The inset in (a) shows a close up of the $\phi \simeq 0$ region.
(c) Zero-phase current $I_J(0)$ as a function of the chemical potential for symmetric ($\mu_L=\mu_R$) and asymmetric ($\mu_L=\mu_0$, with $\mu_0=-1$meV) junctions. $I_J(0)$ is normalized to the critical current $I_c^0$ in the symmetric case with $\mu_L=\mu_R=\mu_0$ and $\tau = 0.8$. Red dashed lines mark the position of the VHSs and color arrows show the situation for the CPRs in the panels above. The inset shows the evolution of $I_J(0)$ as a function of the sublattice symmetry breaking potential $\delta_S$ in the symmetric junction. 
}
\label{fig4}
\end{figure}

Finally, we analyze ZZ junctions in Fig.~\ref{fig4}. 
As discussed above, trigonal warping leads to compensated valley currents at zero phase bias. However, when sublattice symmetry is broken, the nontrivial topology of MATBG captured by the 6BM leads to uncompensated currents and thus $\phi_0$ behavior. This behavior is not observed in the trivial 2BM \cite{SM}. Similarly to AC junctions, we find larger $|I_J(0)/I_c|$ values for asymmetric junctions of n-p type. We also observe a change in sign of $I_J(0)$ through the p-p' to p-n transition. The size of $|I_J(0)/I_c|$ also increases as the sublattice symmetry breaking parameter $\delta_S$ value is increased [inset of Fig.~\ref{fig4}(c)]. It is worth mentioning that we observe asymmetries between the negative and positive critical currents (superconducting diode effect~\cite{daido2022, yuan2022, he2022}) in both AC and ZZ orientations for asymmetric $SS'$ and $SNS$ junctions. 
This effect is observed when the $\phi_0$ behaviour is enhanced and could thus be electrically controlled by tuning the chemical potential and effective transmission. 
Let us further comment that if, additionally, valley polarization is included, e.g., by means of TRS breaking in the parent state as in Ref.~\cite{Xie}, one would observe $\phi_0$ behavior for all orientations.

{\it Minimal model.---} A scattering theory that confirms and gives further insights to the previous results can be built by linearizing the bulk Hamiltonian with respect to $k_{\perp}$ around each Fermi point for a given $k_{\parallel}$~\cite{SM}. For the 6BM, Andreev reflection coefficients for electron to hole processes (and their time reversed) acquire different phases $\theta(k_{\parallel})$ and $\theta'(k_{\parallel})$. Applying matching conditions at the interface corresponding to a single channel junction with transmission 
$T_k=T(k_{\parallel})$, leads to the following set of Andreev bound states~\cite{SM}
\be 
\frac{\epsilon(k_{\parallel})}{\tilde{\Delta}(k_{\parallel})} = \pm \sqrt{T_k+R_k\cos^2 \left(\frac{\bar{\theta}_k}{2} \right) - T_k\sin^2 \left(\frac{\phi}{2}- \frac{\delta\theta_k}{2}\right)} ,
\ee
where $R_k = 1-T_k$, $\tilde{\Delta}(k_{\parallel})$ is an effective gap, $\bar{\theta}_k= \theta(k_{\parallel}) + \theta'(k_{\parallel})$ and $\delta \theta_k = \theta(k_{\parallel})-\theta'(k_{\parallel})$. This simple dispersion relation is valid provided that $T_k < 1$. For the AC junction we get $\theta^K(k_{\parallel}) = \theta^{K'}(-k_{\parallel})$ (same for $\theta'$), which yields that $I_J^K(0) = I_J^{K'}(0)$, in agreement with the full numerical results.
By contrast, the warping distortion of the Fermi surface in the ZZ case results in $\theta(k_{\parallel}) \ne \theta'(k_{\parallel})$, even for the non-topological 2BM \cite{SM}. 
However, in this orientation we find that $\theta^K(k_{\parallel}) = -\theta^{K'}(-k_{\parallel})$ (same for $\theta'$) and, consequently, $I_J^K(0) = -I_J^{K'}(0)$. The ZZ valley currents are thus compensated unless sublattice symmetry is broken and fragile topology taken into account. 

{\it Conclusions:} We have shown that chiral pairing in MATBG junctions would manifest in the appearance of a nonmagnetic $\phi_0$ behavior, stemming from the nontrivial topology of the MATBG bands. This effect can then be used to distinguish chiral superconductivity from other mechanisms such as nodal pairing. Moreover, we illustrated how the effect is enhanced for extended junctions and controlled by electrostatic gating of a middle heavily doped normal region. Although we have focused on $d+id$ case, other chiral symmetries like $p+ip$ or combinations of these symmetries with $s$-wave pairing, would show similar behavior. The orientation sensitivity of the $\phi_0$ effect could help to distinguish this different orbital character in an actual experiment. Furthermore, it would allow us to distinguishing whether the anomalous behavior is due to a broken valley symmetry parent state or due to configurations involving symmetric contributions from both valleys and chiral pairing.

\acknowledgments
We thank F.S. Bergeret, R. Egger, and W.J. Herrera for valuable comments and discussions. 
This work was supported by the Agencia Estatal de Investigaci\'on project No.~PID2020-117992GA-I00 and No.~PID2020-117671GB-I00, the Spanish CM ``Talento Program'' project No.~2019-T1/IND-14088, and through the ``Mar\'{\i}a de Maeztu'' Programme for Units of Excellence in R\&D (CEX2018-000805-M). 


\bibliographystyle{apsrev4-2}

\providecommand{\noopsort}[1]{}\providecommand{\singleletter}[1]{#1}%

\pagebreak
\clearpage


\setcounter{equation}{0}
\renewcommand{\theequation}{S\,\arabic{equation}}

\setcounter{figure}{0}
\renewcommand{\thefigure}{S\,\arabic{figure}}

\onecolumngrid
\vspace{\columnsep}
\section*{Supplemental material to ``Intrinsic non-magnetic \texorpdfstring{$\phi_0$}{TEXT} Josephson junctions in twisted bilayer graphene"}
\vspace{\columnsep}
\twocolumngrid

\section{Description of different models for MATBG} \label{App_Models}

In the main text we have used two different tight-binding models for describing bulk magic angle twisted bilayer graphene (MATBG): the topologically trivial two-band model (2BM) from Ref.~\onlinecite{Koshino2018} and the six-band model (6BM) from Ref.~\onlinecite{Vishwanath2019} that accounts for the MATBG fragile topology. The implementation of these models for normal transport calculations has already been discussed by us in Ref.~\onlinecite{Alvarado2021}. Here, we describe the necessary extensions of these models to account for superconducting transport.   

\begin{figure}[t!]
\centering
\includegraphics[width=8.6cm]{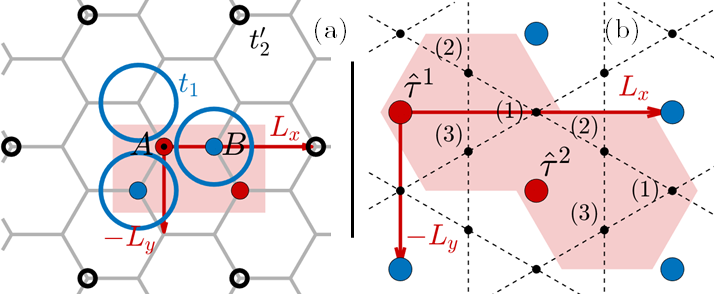}
\caption{Cell doubling real space diagram for (a) 2BM and (b) 6BM, showing the local orbitals and lattice vectors. The shaded red regions in both panels correspond to the doubling of the original triangular primitive cell into a rectangular one. (a) 2BM unit cell and hopping terms associated to the $A^1$ site marked with a black dot in the center. (b) 6BM unit cell where the $p$-orbitals are centered at the red dots $(\hat{\tau}^1,\hat{\tau}^2)$ forming a triangular lattice. The black dots correspond to the $s$-orbitals placed on a Kagome lattice with three sites per minimal unit cell. 
}
\label{fig1_SM}
\end{figure}

In order to obtain the transport properties in defined armchair (AC) and zigzag (ZZ) orthogonal boundaries we have to consider the cell doubling of the unit cell~\cite{Alvarado2022} with lattice vectors $L_x = \sqrt{3} L_m$ and $L_y = L_m$, where $L_m = a/(2\sin{\theta/2}) \approx 13$ nm is the moir\'e lattice length given in terms of the graphene lattice parameter $a$ and the twist angle $\theta$, see Fig.~\ref{fig1_SM}. Local fermion operators in the whole moir\'e superlattice Hilbert space take the form $\hat{\Psi} = (\hat{\Psi}_\alpha^1 \; \hat{\Psi}_\alpha^2)^T$, where $\alpha$ accounts for the rest of orbital degrees of freedom. The Hamiltonian of the normal state in the cell-doubled space adopts the general form 
\be \label{general_cd}
\hat{H}_e = \bmat \hat{H}^{11} && \hat{H}^{12} \\
\hat{H}^{12 \dagger} && \hat{H}^{22} \emat.
\ee

Concerning the implementation of superconductivity in MATBG, we have focused on configurations involving symmetric contributions from both valleys, which are the most likely ones according to many previous theoretical studies for the SC domes around filling $\nu = -2$ (see Refs.~\cite{po2018, shavit2021, lake2022}). Therefore, we exclude further effects coming from possible flavor resets when varying the band filling of the flat bands. 
Regarding the rest of degrees of freedom of the system, we adopt the rather natural spin projection $S^z=0$ intervalley chiral superconductivity with zero net momentum Cooper pairs~\cite{khalaf2022}. 
Despite the fact that the orbital character of the pairing cannot be determined by existent experimental data, we will focus on $d+id^\prime$ wave as a typical example of time reversal symmetry (TRS) breaking topological superconductivity. In any case, $p+ip^\prime$ with $S^z=0$ spin projection shows similar properties to the chiral $d$-wave, featuring the aforementioned $\phi_0$ behaviour (not shown in this work). 

We thus characterize the bulk MATBG by a Bogoliubov-de Gennes Hamiltonian (BdG),
\begin{equation}
\label{HBDG}
    \hat{\mathcal{H}}(\vec{k}) = \left(\begin{array}{cc} \hat{H}_e(\vec{k})-\mu \hat{\mathbb{I}} & \hat{\Delta}(\vec{k}) \\
   \hat{\Delta}^{\dagger}(\vec{k}) & \mu \hat{\mathbb{I}} - \hat{H}_h(\vec{k}) \end{array} \right) ,
\end{equation}
where the electron (hole) Hamiltonian $\hat{H}_e(\vec{k})$ [$\hat{H}_h(\vec{k})$] and pairing $\hat{\Delta}(\vec{k})$ can take into account the multiband, fragile topology~\cite{Vishwanath2019} structure of MATBG. For an intervalley implementation of the pairing on the $K$ valley we use $\hat{H}_h(\vec{k}) = \mathcal{T}\hat{H}^K_e(\vec{k}) \mathcal{T}^\dagger = \hat{H}^{K'}_e(-\vec{k})^* $. Equivalently, for TRS-breaking intravalley pairing we have $\hat{H}_h(\vec{k}) = \mathcal{T}\hat{H}^{K'}_e(\vec{k}) \mathcal{T}^\dagger = \hat{H}^{K}_e(-\vec{k})^*$. The intravalley term is used to rule out nodal, TRS-breaking (non chiral) pairing mechanism as a source of $\phi_0$ behaviour.

\subsection{Superconducting 6BM Hamiltonian with unit cell doubling}

The normal state of the 6BM is defined in a basis of $p_z$ and $p_\pm$-orbitals in a triangular lattice, $(\tau, p_z)$ and $(\tau, p_\pm)$, respectively, and three $s$-orbitals in a Kagome lattice $(\kappa, s)$, see Fig.~\ref{fig1_SM}(a). The local fermion operators in the doubled unit cell are defined as 
$\hat{\Psi}^{\mu}_\alpha =
(\hat{\tau}_{p_z, \mu} \; \hat{\tau}_{p_+, \mu} \; \hat{\tau}_{p_-, \mu} \; \hat{\kappa}^{(1)}_{s, \mu}\; \hat{\kappa}^{(2)}_{s, \mu}\; \hat{\kappa}^{(3)}_{s, \mu}\;)^T$,
where $\mu \equiv 1,2$ indicates the two moir\'e sites within the orthogonal cell. The general structure of the Hamiltonian follows
\bea
&\hat{H}^{11} = \bmat H_{p_z}^{11} +\mu_{p_z} & \hat{C}_{p_\pm p_z}^{11} & \hat{0} \\
        \hat{C}_{p_\pm p_z}^{11 \dagger} & \hat{H}_{p_\pm}^{11}+\mu_{p_\pm}\hat{\mathbb{I}}_2 & \hat{C}_{\kappa p_\pm}^{11} \\
        \hat{0} & \hat{C}_{\kappa p_\pm}^{11 \dagger} & \hat{H}_{\kappa}^{11} +\mu_\kappa\hat{\mathbb{I}}_3 \emat, \nonumber \\
&\hat{H}^{12} = \bmat H_{p_z}^{12} & \hat{C}_{p_\pm p_z, 1}^{12} & \hat{0} \\
        \hat{C}_{p_\pm p_z, 2}^{12} & \hat{H}_{p_\pm}^{12} & \hat{C}_{\kappa p_\pm, 1}^{12} \\
        \hat{0} & \hat{C}_{\kappa p_\pm, 2}^{12} & \hat{H}_{\kappa}^{12} \emat ,
\eea
where $\hat{\mathbb{I}}_{2(3)}$ is the identity matrix in two (three) dimensions, $\hat{H}^{22} = \hat{H}^{11}$, and the expressions for the matrices $H^{\mu\nu}_{\alpha}$ and $C^{\mu\nu}_{\alpha\beta}$ and the parameters used are given in Ref.~\onlinecite{Alvarado2021}.

The low energy physics associated to the flat bands in the 6BM are described fundamentally by the directional $p_\pm$-orbitals in the triangular lattice $(\tau,p_\pm)$~\cite{Vishwanath2019}. 
Therefore, we project the pairing order parameter over these orbitals only. 
We restrict ourselves to the simplest case in which there is no sublattice structure. 
The pairing wavefunctions that preserve the global symmetry of the lattice are obtained from the irreducible representations of the crystal symmetry group~\cite{pangburn2022} in the triangular lattice, 
\bea
\Delta_{d_{x^2-y^2}} &=& \frac{\Delta}{\sqrt{3}} \left[ \cos{(k_y L_y)}-\cos{(k_x L_x/2)}\cos{(k_y L_y/2)} \right], \nonumber \\
\Delta_{d_{xy}} &=& \Delta\sin{(k_x L_x/2)}\sin{(k_y L_y/2)},
\eea  
Recent experiments~\cite{oh2021} based on Andreev reflection measurements point towards $\Delta \approx 0.3$ meV. In our calculations we set the value $\Delta = 0.1$ meV, one order of magnitude smaller than the bandwidth. 

We project this form factors in the $(\tau,p_\pm)$ subspace expanded to take into account the moir\'e superlattice degree of freedom in the basis $\hat{\Psi}_\pm = (\hat{\Psi}^1_{\pm,\up} \; \hat{\Psi}^2_{\pm,\up} \; \hat{\Psi}^{1\dagger}_{\pm,\dw} \; \hat{\Psi}^{2\dagger}_{\pm,\dw})^T$, with $\hat{\Psi}^{\mu}_{\pm, \sigma} =
(\hat{\tau}^\mu_{p_+, \sigma} \; \hat{\tau}^\mu_{p_-, \sigma})^T$ and $\sigma = \uparrow \downarrow$ being the implicit spin degree of freedom. 
We observe the following intra-orbital contributions for the nodal terms $d = d_{x^2-y^2}$ and $d'=d_{xy}$,
\begin{subequations}
\begin{align}
\hat{\Delta}^{11}_{d,p_\pm} ={}& (\phi_{01}+\phi_{0\bar{1}})/(2\sqrt{3}) \; \hat{\mathbb{I}}_2, 
\\
\hat{\Delta}^{12}_{d,p_\pm} ={}& - (1 + \phi_{0\bar{1}})(1+\phi_{10})/(4\sqrt{3}) \; \hat{\mathbb{I}}_2, 
\\
\hat{\Delta}^{11}_{d',p_\pm} ={}& \hat{0}, 
\\
\hat{\Delta}^{12}_{d',p_\pm} ={}& (1 - \phi_{0\bar{1}})(1-\phi_{10})/4 \; \hat{\mathbb{I}}_2,
\end{align}
\end{subequations}
where $\phi_{10} = e^{-i k_x L_x}$ and $\phi_{01} = e^{-i k_y L_y}$, with $\phi_{\bar{ij}} = \phi_{ij}^\dag$. 

If we explicitly show the total electron and hole contributions of the Hamiltonian in the moir\'e superlattice space we obtain
\be 
\quad \hat{\Delta}_{x,p_\pm} = \bmat \hat{\Delta}^{11}_{x,p_\pm} & \hat{\Delta}^{12}_{x,p_\pm}  \\ \hat{\Delta}^{12 \dagger}_{x,p_\pm}  & \hat{\Delta}^{11}_{x,p_\pm}  \emat,
\ee
where $x = d,d'$ and the total chiral contribution for each valley is $\hat{\Delta}_{p_\pm} = \hat{\Delta}_{d,p_\pm}+i\hat{\Delta}_{d',p_\pm}$.

The normal-state Hamiltonian includes the possibility of breaking the graphene sublattice degree of freedom by means of the perturbation $\hat{V}^{11} = \hat{V}^{22} =\delta_S\hat{\tau}^{p_\pm}_z$, where  $\hat{\tau}^{p_{\pm}}_z$ is a Pauli matrix acting over the $p_{\pm}$ orbitals. Experiments until this date have shown an incompatibility between superconductivity and full $h$BN alignment, but symmetry breaking superconducting domes have been observed close to charge neutrality~\cite{lu2019,oh2021}. For the purpose of this work, graphene's sublattice symmetry breaking can act as an example of neither magnetic, nor valley, polarization mechanism to obtain uncompensated valley currents for the ZZ case. 

\subsection{Superconducting 2BM Hamiltonian with unit cell doubling}

We use the topologically trivial 2BM as a test to compare the features associated primarily to the Fermi surface properties and the ones in which the fragile topology is involved. We use the same pairing wavefunction projected in the triangular lattice, as for the 6BM. However, in the hexagonal lattice of the 2BM, we only consider the next nearest neighbours ($nnn$) terms to expand the order parameter, neglecting the nearest neighbours ($nn$) ones, so we can directly compare the effects of the Fermi surface over the exact same pairing wavefunction. We implement the chiral $d$-wave using the following spinors $\hat{\Psi} = (\hat{\Psi}^1_{\up} \; \hat{\Psi}^2_{\up} \; \hat{\Psi}^{1\dagger}_{\dw} \; \hat{\Psi}^{2\dagger}_{\dw})^T$, with $\hat{\Psi}^{\mu}_{\sigma} =
(\hat{\psi}^\mu_{B, \sigma} \; \hat{\psi}^\mu_{A, \sigma})^T$. We first start with a simplified version of the cell doubled normal Hamiltonian with respect to the one used in Ref.~\onlinecite{Alvarado2021}, just considering $t_1$ and $t_2$ in a slightly different unit cell, see Fig.~\ref{fig1_SM}(b). Particularizing Eq.~\eqref{general_cd} to the 2BM we have
\bea
 \hat{H}^{11} &=& \bmat h_0-\mu & t_1 (1 + \phi_{0\bar{1}}) \\ t_1 (1 +\phi_{01}) & h_0-\mu \emat, \nonumber \\
\hat{H}^{22} &=& \bmat h_0-\mu & t_1 (1 + \phi_{01}) \\ t_1 (1 +\phi_{0\bar{1}}) & h_0-\mu \emat, \nonumber \\
\hat{H}^{12} &=& \bmat h_1 & t_1 \phi_{\bar{1}0} \\ 
t_1 & h_2 \emat, 
\eea
where
\begin{subequations}
\begin{align}
h_0 ={}& t_2 \phi_{10} + t^*_2 \phi_{\bar{1}0}, 
\\
h_1 ={}& \phi_{01} (t^*_2 + t_2 \phi_{\bar{1}0}) + \phi_{0\bar{1}}^2 (t^*_2 + t_2 \phi_{\bar{1}0}), 
\\
h_2 ={}& \phi_{0\bar{1}} (t^*_2  + t_2 \phi_{\bar{1}0}) + \phi_{01}^2 (t^*_2 + t_2 \phi_{\bar{1}0}).
\end{align}
\end{subequations}
Finally we can break sublattice symmetry adding the perturbation $\hat{V}^{11} = \hat{V}^{22} = \delta_S\hat{\tau}^{AB}_z$, where $\hat{\tau}$ is a Pauli matrix.

The pairing contributions for the nodal $d$ and $d'$ have the form
\begin{subequations}
\begin{align}
\hat{\Delta}^{11}_{d} ={}& (\phi_{01}+\phi_{0\bar{1}})/(2\sqrt{3}) \; \hat{\mathbb{I}}_2, 
\\
\hat{\Delta}^{12}_{d,AA} ={}& - (1 + \phi_{01})(1+\phi_{\bar{1}0})/(4\sqrt{3}), \;  
\\
\hat{\Delta}^{12}_{d,BB} ={}& - (1 + \phi_{0\bar{1}})(1+\phi_{\bar{1}0})/(4\sqrt{3}), 
\\
\hat{\Delta}^{11}_{d'} ={}& \hat{0}, 
\\
\hat{\Delta}^{12}_{d',AA} ={}& (1 - \phi_{01})(1-\phi_{\bar{1}0})/4,  
\\
\hat{\Delta}^{12}_{d',BB} ={}& -(1 - \phi_{0\bar{1}})(1-\phi_{\bar{1}0})/4.
\end{align}
\end{subequations}

Finally the chiral order parameter can be expressed as
\be
\hat{\Delta}^{12}_{x} = \bmat \hat{\Delta}^{12}_{x,BB} & 0  \\ 0  & \hat{\Delta}^{12}_{x,AA}\emat, 
\; \hat{\Delta}_{x} = \bmat \hat{\Delta}^{11}_{x} & \hat{\Delta}^{12}_{x}  \\ \hat{\Delta}^{12 \dagger}_{x}  & \hat{\Delta}^{11}_{x}\emat, 
\ee
where $x = d,d'$ and $\hat{\Delta} = \hat{\Delta}_{d}+i\hat{\Delta}_{d'}$.

\subsection{Analysis of superfluid response}

To observe the inherited topological effects in the pairing, we now analyze the superfluid weight. The superfluid weight gives the size of the superfluid current for a given phase gradient in the bulk, and it is connected to central phenomenology of the superconducting phase, like the Meissner effect. 

Recent works relate the superfluid weight in multiband systems featuring superconducting flat bands with topological properties like the Berry curvature~\cite{peotta2015, liang2017, hu2019, julku2020, torma2021, rossi2021}.  
Following Ref.~\onlinecite{julku2020}, we study the superfluid weight in the doubled unit cell defined via the static Meissner effect for local and nonlocal interactions
\begin{widetext}
\bea
D^s_{\mu\nu} = \frac{1}{V} \sum_{k,n,m} \frac{f(E^m_k)-f(E^n_k)}{E^n_k-E^m_k} && ( \bra\Psi^n_k|\partial_\mu \hat{\mathcal{H}}_k(\Delta=0) | \Psi^m_k \ket  \bra\Psi^m_k|\partial_\nu \hat{\mathcal{H}}_k | \Psi^n_k \ket - \nonumber \\
&&\bra\Psi^n_k|\partial_\mu \hat{\mathcal{H}}_k(\Delta=0) \hat{\tau}_z| \Psi^m_k \ket  \bra\Psi^m_k|\partial_\nu \hat{\tau}_z \hat{\mathcal{H}}_k(\Delta=0) | \Psi^n_k \ket ),
\eea
\end{widetext}
where $\mu,\nu=\{x,y\}$ are the different components of the superfluid weight tensor marking the orientation of the partial derivatives of the momentum $\partial_\mu = \partial_{k_\mu}$, $V$ is the area of the sample in k-space, $f(E)$ is the Fermi-Dirac distribution in the zero temperature limit, and $\hat{\cal{H}}|\Psi^{n}_k\ket=E^{n}_k|\Psi^{n}_k\ket$ represent the eigenvalue problem for the $n$-th band of the BdG Hamiltonian at momentum $k$ contained in the first Brillouin zone. 

\begin{figure}[t!]
\centering
\includegraphics[width=8.6cm]{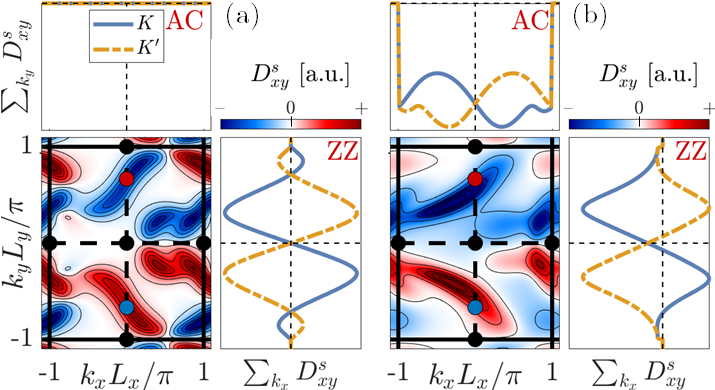}
\caption{Unconventional anisotropic pairing behaviour from superfluid weight calculations showing $D^s_{xy}\neq0$ even for local $s$-wave pairing. Main panels show trivial 2BM (a) and the fragile-topological 6BM (b) for the $K$ moir\'e valley at band fillings showing typical Fermi surfaces. The solid (dot-dashed) line in the lateral panels represents the integrated superfluid weight in the $k_\perp$ direction for the AC and ZZ orientations for $K$ ($K'$) valley. The 6BM shows an uncompensated $D^s_{xy}$ component, specially evident for the AC orientation. For visualization purposes the pairing potential in 2BM (6BM) case is $\Delta = 0.4$ ($\Delta = 1$) meV. }
\label{fig2_SM}
\end{figure}

The $D^s_{xy}$ component of the superfluid weight, Fig.~\ref{fig2_SM}, shows nonzero values for the 6BM case. This feature is associated to spontaneous $C_3$ rotational symmetry breaking and thus unexpected nematic response for the 6BM, which is specially noticeable for the AC orientation. Notice that this effect is not compensated between valleys in contrast to the 2BM ZZ case. Previous microscopic tight-binding calculations of the superfluid weight showed anisotropies in conventional MATBG bulk when the pairing is nonlocal~\cite{julku2020}. This phenomenon can be explained by the nontrivial multi-orbital character of the 6BM that is captured in the quantum geometric tensor and, consequently, in the superfluid weight~\cite{torma2021}. The value of the anisotropic response is proportional to $\Delta$ and equivalent for local and nonlocal pairing wavefunctions. 

The anisotropic $D_{xy}$ component demonstrates the presence of inherited unconventional properties associated to the pairing in the 6BM bulk when the doubled unit cell is considered. This anomalous response is specially guaranteed for the AC orientation and exemplifies the different properties of the 2BM and 6BM even in the superconducting phase, despite the fact that both models have the same Chern number when chiral superconductivity is taken into account. 

\section{Transport Calculations} \label{App_Scattering}
The phase-biased equilibrium Josephson current can be obtained using Keldysh Green functions (GFs)~\cite{Zazunov2016} as
\be
I = \frac{e}{h}  \int \frac{d\omega \, dk_\parallel}{\Omega_{k_\parallel}} \, \textrm{tr}_N \left \{ \hat{\tau}_z (\hat{\Sigma}_{LR} \hat{G}_{RL}^{+-} - \hat{\Sigma}_{RL} \hat{G}_{LR}^{+-} )\right \},
\ee 
where $\Omega_{k_\parallel} = 2\pi/L_\parallel$ accounts for the integration limits, the trace tr$_N$ and Pauli matrix $\hat{\tau}_z$ act over Nambu space, $\hat{\Sigma}_{LR}$ is the coupling between the leads and $\hat{G}^{+-}$ are the Keldish components of the junction GF. At equilibrium we have
\be 
\hat{G}_{j j^\prime}^{+-} = n_F(\omega) [\hat{G}_{j j^\prime}^A-\hat{G}_{j j^\prime}^R] = n_F(\omega) [\hat{G}_{j j^\prime}^A-(\hat{G}_{j^\prime j}^A)^\dag],
\ee 
where $j,j^\prime = L,R$ are written in terms of the advanced GF (from now on we omit the superindex $A$). We obtain the edge GFs $\hat{G}_{j j^\prime}$ from the full junction GF
\bea
&\check{\hat{G}} = 
\bmat 
\hat{G}_{LL} && \hat{G}_{LR} \\
\hat{G}_{RL} && \hat{G}_{RR}
\emat =
\bmat 
\hat{\cal G}_L^{-1} && -\hat{\Sigma}_{LR} \\
-\hat{\Sigma}_{RL} && \hat{\cal G}_R^{-1}
\emat^{-1}, & \nonumber \\ \nonumber \\ 
&\hat{G}_{LR} = \hat{\cal G}_L\hat{\Sigma}_{LR} \hat{G}_{RR},& \nonumber \\ 
&\hat{G}_{RR} = [\hat{\mathbb{I}} -  \hat{\cal G}_R \hat{\Sigma}_{LR}^\dag \hat{\cal G}_L \hat{\Sigma}_{LR}]^{-1}\hat{\cal G}_R,&
\eea  
where $\hat{\cal G}_{L/R}$ is the boundary GF of the left (right) side associated to the superconducting MATBG lead. To compute these boundary GFs, we use the recursive method described in Ref.~\onlinecite{Alvarado2022} (notice that in the doubled cell space the 6BM is described up to $nn$ and the 2BM up to $nnn$). 

Experimental setups typically consist on a TBG slab with several electrodes but no constrictions or abrupt junctions (e.g., other metal oxides, etc.). Our description must recover a perfect bulk when no phase bias is applied and the chemical potential along the sample is homogeneous. To do so, we use the following general self-energy structure for the junctions
\be 
\hat{\Sigma}_{LR}(k_\parallel, \phi)  = \tau \hat{T}_{LR}(k_\parallel) \hat{\tau}_z e^{i\hat{\tau}_z \phi/2},
\ee
where $\tau$ is an effective transmission coefficient of the junction and $\hat{T}_{LR}(k_\parallel)$ are the nonlocal contributions of the BdG Hamiltonian such that for a symmetric junction and $\tau=1$ we recover a pristine bulk. 
With this general structure, and taking advantage of the recursive Green function method, we define three different types of junction. First, a symmetric $SS$ junction, in which $\hat{\mathcal{H}}_L = \hat{\mathcal{H}}_R$, defined by a phase bias and a non-perfect transmission $\tau<1$. Equivalently, but varying the filling on one of the leads with respect to the other, we obtain an asymmetric $SS^\prime$ junction where $\mu_L \neq \mu_R$. Finally, we can compute the more realistic junction in which $\tau=1$ and the phase bias is applied along a finite normal region of a few moir\'e wavelengths. This corresponds to a $SNS$ junction in which the normal region is heavily doped $|\mu_L| = |\mu_R| \ll |\mu_c|$. 

\begin{figure}[t!]
\centering
\includegraphics[width=8.6cm]{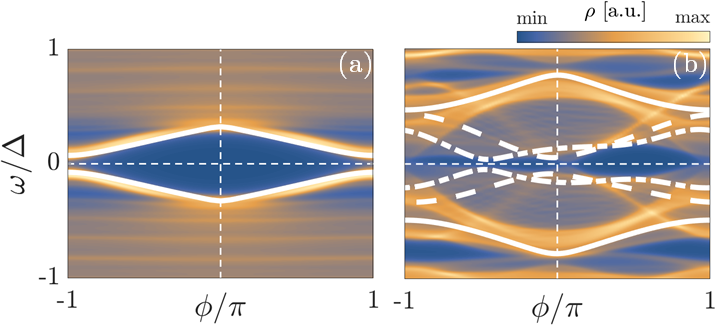}
\caption{Spectral density $\rho(E,\phi)$ at the $K$ moir\'e valley for AC junctions based on the trivial 2BM (a) and the topological 6BM (b). We set $\tau=0.8$ and use $\mu_L=\mu_R=-0.25$ meV for 2BM and $\mu_L=\mu_R=-1$ meV for the 6BM. White solid, dot-dashed, and dashed lines represent the lowest energy ABSs for the discrete set of parallel momenta $k_\parallel = [\bar{\Gamma}, \bar{X}/2, \bar{X}]$, respectively.}
\label{fig3_SM}
\end{figure}

In Fig.~\ref{fig3_SM} we show results for the subgap Andreev spectrum $\rho(E,\phi)$ obtained for the two different models for the case of a $SS$ junction. The subgap spectrum is dramatically different for the two models: 
it satisfies $\rho(E,\phi) = \rho(E,\pi-\phi)$ for the 2BM while this symmetry is broken in the 6BM. 
In particular, the phase shift ($\phi_0$) for Andreev bound states (ABSs) in the 6BM varies with $k_\parallel$, featuring maximum phase displacement at $\bar{X}/2$. The ABSs satisfy that $\epsilon^K(k_\parallel,\phi) = \epsilon^{K'}(-k_\parallel,\phi)$, enabling the appearance of a net $\phi_0$ behaviour despite the different contributions from each transport channel at each valley. 

\subsection{Linearized scattering approach}

To obtain the CPR from a scattering perspective, one needs to determine the Andreev reflection coefficients at an ideal interface between a normal and a superconducting MATBG region, assuming conservation of the momentum parallel to the junction interface $k_{\parallel}$. We start by linearizing Eq. (\ref{HBDG}) around the $i$-th Fermi point with respect to the incident momentum $k^i_{\perp}$, i.e.,
\bea
&\hat{\mathcal{H}}^i \approx \hat{H}^i_0 + \hat{H}^i_1\delta k^i_\perp + O\left [(\delta k^{i}_\perp)^2 \right ],& \nonumber \\
&\hat{H}^i_n = \frac{1}{n!} \left.\frac{\partial^n \hat{\mathcal{H}}}{\partial k_\perp^n}\right|_{k_\perp^i}. &
\eea
The scattering states $|\Psi^i_E\rangle$ would then satisfy the equation for the correction of the momentum with respect to the Fermi point, $\delta k^{i}_\perp$, 
\begin{equation}
\label{scattering}
\hat{\mathcal{A}}^{-1}\left(\hat{H}^i_0-E\hat{\mathbb{I}}\right)|\Psi^i_E\rangle = \delta k^i_\perp(E) |\Psi^i(E)\rangle,
\end{equation}
where $\hat{\cal{A}}$ is the adjoint matrix of $\hat{H}^i_1$.

We can further simplify the problem by projecting Eq.~(\ref{scattering}) into the subspace spanned by the states $|\Psi^i_{\nu;\alpha}\rangle$, corresponding to the dominant electron ($\nu=e$) and hole ($\nu=h$) states at the $i$-th Fermi point. The label $\alpha$ denotes additional degrees of freedom which could be coupled by scattering at the junction interface. Finally, we project the problem into a $2 \times 2$ Nambu space, namely,
\begin{subequations}
\begin{align}
\langle \Psi^i_e| \hat{\mathcal{A}}^{-1} \left(\hat{H}^i_0-E\hat{\mathbb{I}}\right) |\Psi^i_e \rangle  ={}& v_F E - \partial_{k} \Delta \Delta^\dag, 
\\
\langle \Psi^i_e| \hat{\mathcal{A}}^{-1} \left(\hat{H}^i_0-E\hat{\mathbb{I}}\right) |\Psi^i_h \rangle  ={}& -v_F \Delta + \partial_{k} \Delta E, 
\\
\langle \Psi^i_h| \hat{\mathcal{A}}^{-1} \left(\hat{H}^i_0-E\hat{\mathbb{I}}\right) |\Psi^i_e \rangle  ={}& v_F \Delta^\dag + \partial_{k} \Delta^\dag E, 
\\
\langle \Psi^i_h| \hat{\mathcal{A}}^{-1} \left(\hat{H}^i_0-E\hat{\mathbb{I}}\right) |\Psi^i_h \rangle  =& -v_F E - \partial_{k} \Delta^\dag \Delta, 
\end{align}
\end{subequations}
where $E$ is the incident energy of the quasiparticle and $\partial_{k} = \partial_{k_\perp}$. The equation for $\delta k^{i}_\perp$ acquires the eigenvalue problem structure
\be
\bmat
- v_F & - \partial_k \Delta \\
- \partial_k \Delta^\dag & v_F
\emat \bmat 
-E & \Delta \\ \Delta^\dag & -E
\emat 
|\Psi^i_E\rangle =  -\lambda|\Psi^i_E\rangle,
\ee
with $\delta k^i_\perp(E) = \lambda/\det \hat{\mathcal{A}}_i$. The eigenvalues of $\delta k^i_\perp(E)$, that is, the corrections to the Fermi momentum in the direction perpendicular to the junction, are
\be
\lambda = \alpha \pm \sqrt{(\beta- v_F E)^2 - (v_F \Delta - \partial_{k} \Delta E)(v_F \Delta^\dag + \partial_{k} \Delta^\dag E)},
\ee
with 
\be
\alpha = \frac{\partial_{k} \Delta \Delta^\dag+\partial_{k} \Delta^\dag\Delta}{2},\quad \beta = \frac{\partial_{k} \Delta \Delta^\dag -\partial_{k} \Delta^\dag\Delta}{2}.
\ee
The eigenvalues $\delta k_{\pm}^i(E)$ (with associated eigenstates $|\Psi^i_{\pm}(E)\rangle$) become complex at subgap energies. Consequently, we can extract two Andreev reflection coefficients for each Fermi point, $z^i_{\pm}(E) = \langle \Psi^i_{h}|\Psi^i_{\pm}(E)\rangle/\langle \Psi^i_{e}|\Psi^i_{\pm}(E)\rangle$.

\begin{figure}[t!]
\centering
\includegraphics[width=8.8cm]{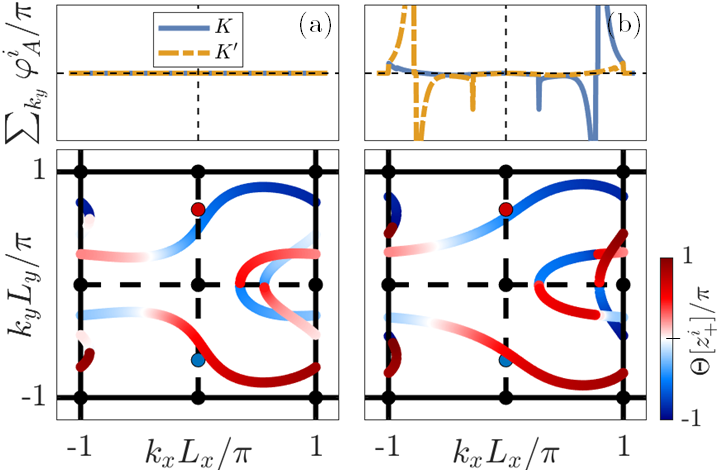}
\caption{Andreev phase of $z^i_+=u_+^i/v_+^i$ based on the trivial 2BM (a) and 6BM (b) models for the $K$ moir\'e valley for band filling showing typical Fermi surfaces. The solid (dot-dashed) line in upper panels represents the integrated Andreev phase in the $k_\perp$ direction for the AC direction for the $K$ ($K'$) valley. For visualization purposes the pairing potential in 2BM (6BM) case is $\Delta = 0.4$ ($\Delta = 1$) meV.}
\label{fig4_SM}
\end{figure}

While for a conventional single-band superconductor the Andreev coefficients on opposite Fermi points
exhibit exactly opposite phases, for general MATBG models with chiral pairing this inversion symmetry can be broken, even when it is preserved in the normal state as in the case of the AC orientation. The 2BM model satisfies that the eigenvalues $\delta k^i_\perp(E)$ for the chiral $d$-wave come in complex conjugated pairs and thus we find that $\partial_{k} \Delta \Delta^\dag = (\partial_{k} \Delta^\dag \Delta )^\dag$ for $E=0$. 
The nontrivial behavior of the 6BM is revealed in the phase of the Andreev reflection coefficients obtained from the eigenvectors $z^i_{\pm} = u^i_{\pm}/v^i_{\pm}$, see Fig.~\ref{fig4_SM}. 
The phase accumulated along the Fermi surface in the AC (perpendicular) direction is zero for the 2BM and nonzero for the 6BM. The valley degree of freedom does not alter this net behaviour. 
We note that the Andreev phase analysis for intravalley pairing shows uncompensated valley phases (i.e., signatures of $\phi_0$ behaviour) \textit{only for chiral pairings}, despite the TRS-breaking nature of all intravalley pairings, including the nodal ones. 

In general, the $z^i_{\pm}$ coefficients on opposite Fermi points satisfy the following approximate ansatz 
for AC junctions based on the previous numerical results,
\bea 
z^1_{+}(E) &=& X^1_+(E)e^{i\theta_k} , \;\;\, \quad z^1_{-}(E) = X^1_-(E)e^{i\theta'_k},\nonumber \\
z^2_{+}(E) &=& X^2_+(E)e^{-i\theta_k}, \quad z^{2}_{-}(E) = X^2_-(E)e^{-i\theta'_k}, \nonumber \\
\eea 
with 
\bea \label{eq:ansatz-X}
&X_{\pm}^2(E) \simeq (E \mp i \rho^{\pm 1}\sqrt{|\tilde{\Delta}_k|^2-E^2})/|\tilde{\Delta}_k|,& \nonumber \\
&X_{\pm}^1(E) = 1/(X_{\pm}^2(E))^*,&
\eea
where $\tilde{\Delta}_k$ is an effective gap which depends on $k_{\parallel}$ and $\rho \gtrsim 1$ a dimensionless parameter. In the AC configuration we have two symmetric Fermi points $k_\perp^{1,2} \equiv \mp k_F$. 

The boundary modes take the form
\bea 
&\Psi_L^+ = \bmat 1 \\ X_-^2e^{-i \theta_k^\prime} \emat, \quad
\Psi_L^- = \bmat 1 \\ X_+^1e^{i \theta_k} \emat,& \nonumber \\
&\Psi_R^+ = \bmat 1 \\ X_+^2e^{-i \theta_k} \emat, \quad
\Psi_R^- = \bmat 1 \\ X_-^1e^{i \theta_k^\prime} \emat.& 
\eea 
Notice that right and left moving solutions have a different phase of the Andreev coefficients for each Fermi point, $\theta_k = \theta(k_\parallel)$ and $\theta'_k = \theta'(k_\parallel)$, thus breaking inversion symmetry.

The boundary condition at the interface, imposing a single channel matching with effective transmission $T(k_\parallel) = T_k$ and fulfilling $R_k+T_k=1$, takes the form
\bea   
&\tau_k a \Psi^+_L =  e^{i \sigma_z \phi/2} \left(c \Psi^+_R + r_kd \Psi^-_R   \right)&, \nonumber \\
&\tau_k b \Psi^-_L = e^{i \sigma_z \phi/2} \left(r_kc \Psi^+_R + d \Psi^-_R   \right)&,
\eea
where $\tau_k = \sqrt{T_k}$, $r_k = \sqrt{R_k}$, and $\sigma_z$ is a Pauli matrix acting in Nambu space. 
The equation for the Andreev bound states is thus obtained imposing that the determinant of the matrix of coefficients ($a$, $b$, $c$, $d$) is zero. 
In general, the solutions of this equation, $E_{0}$, are complex numbers 
\be
E_0(k_{\parallel}) \simeq \epsilon(k_{\parallel}) + i \Gamma(k_{\parallel}), 
\ee
with both $\epsilon(k_{\parallel})$ and $\Gamma(k_{\parallel})$ real. 
The energy (real part) of the bound state equation adopts the simple form
\be \label{eq:ABS-re}
\frac{\epsilon(k_{\parallel})}{\tilde{\Delta}(k_{\parallel})} \simeq \pm \sqrt{T_k+R_k\cos^2 \left(\frac{\bar{\theta}_k}{2} \right) - T_k\sin^2 \left(\frac{\phi}{2}- \frac{\delta\theta_k}{2}\right)},
\ee
with $\bar{\theta}_k= \theta_k + \theta'_k$ and $\delta \theta_k = \theta_k-\theta'_k$. 
But the imaginary part $\Gamma(k_{\parallel})$ is, in general, nonzero. 
For $T_k < 1$, however, we find that the imaginary part vanishes and we have real Andreev bound states in the limit $\rho \rightarrow 1$. 
Equation (\ref{eq:ABS-re}) thus shows that the $\phi_0$ effect originates from the symmetry breaking induced by the different phases obtained from the eigenvalue problem in the 6BM case. 

For the AC junction, the 2BM is recovered when $\theta_k^\prime = \theta_k$ so the phase shift vanishes, $\delta \theta_k=0$. In this case, the trivial phases accumulated only produce a renormalization of the effective transmission of the junction.
In the AC 6BM, we observe that $\theta^K(k_{\parallel}) = \theta^{K'}(-k_{\parallel})$ and, equivalently, for $\theta'$ (see Fig.~\ref{fig4_SM}). 
Consequently, we observe different $\phi_0$ contributions for each $k_\parallel$ that integrate to the same $\phi_0$-shifted total valley currents with equal magnitude and sign. 

For ZZ junctions, in both models we observe compensated valley currents $I_J^K(0) = -I_J^{K'}(0)$ for $E\neq0$, in agreement with the full numerical results. 
Only in the 6BM, and when sublattice symmetry is broken, we expect to observe uncompensated valley currents due to asymmetries between valleys induced in the eigenvalues of the scattering problem.

Finally, when intravalley pairing is considered, we only observe $\phi_0$ behaviour for chiral symmetries and under the same conditions as studied in the intervalley case, in agreement with the full numerical results. By contrast, we observe compensated valley currents for asymmetric $SS'$ junctions when nodal pairing is studied,
as predicted by the accumulated phases in the Andreev coefficients analysis. These valley currents for both AC and ZZ orientations are always compensated $I_J^K(0) = -I_J^{K'}(0)$, even when graphene's sublattice symmetry is broken. Consequently, in the case of nodal pairing other valley polarizing mechanism would be required to lift the symmetry and observe the $\phi_0$ effect. 

Therefore, both the scattering and the full numerical analysis link the anomalous Josephson effect only to chiral pairing in MATBG. No other TRS-breaking symmetry, like nodal intravalley, would exhibit $\phi_0$ behaviour in direct Josephson junctions. 
As we explain in the main text, this fact provides a direct way to distinguish between nodal and full-gapped chiral pairing in MATBG.

\end{document}